\documentclass{emulateapj}

\newcommand{\xte}{{\it RXTE}}
\newcommand{\eps}{{\rm ergs\,s^{-1}}}
\newcommand{\epcs}{{\rm ergs\,cm^{-2}\,s^{-1}}}
\newcommand{\cts}{{\rm count\,s^{-1}}}
\newcommand{\src}{4U~1901+03}

\slugcomment{Accepted by \apj}

\shorttitle{Discovery of pulsations in \src}
\shortauthors{Galloway et al.}

\begin{document}

\title{Discovery of pulsations in the X-ray transient 4U~1901+03}

\author{Duncan K. Galloway\altaffilmark{1}, 
   Zhongxiang Wang,
and
   Edward H. Morgan}
\affil{ Kavli Institute for Astrophysics and Space Research,\\
   Massachusetts Institute of Technology, Cambridge, MA 02139}

\email{D.Galloway@physics.unimelb.edu.au, wangzx@space.mit.edu,
  ehm@space.mit.edu}

\altaffiltext{1}{present address: School of Physics, University of
Melbourne, Victoria 3010, Australia}

\begin{abstract}
We describe observations of the 2003 outburst of 
the hard-spectrum X-ray transient
\src\ with the {\it Rossi X-ray Timing Explorer}. 
The outburst was first detected in
2003 February by the All-Sky Monitor,
and reached a peak 2.5--25~keV flux of 
$8\times10^{-9}\ \epcs$ (around 240~mCrab).
The only other known outburst occurred 32.2~yr earlier, likely the longest
presently known recurrence time for any X-ray
transient.
Proportional Counter Array (PCA) observations 
over the 5-month duration of the 2003 outburst revealed a
2.763~s pulsar in a
22.58~d orbit.
The detection of pulsations down to a
flux of $3\times10^{-11}\ \epcs$ (2.5--25~keV), along with the inferred
long-term accretion rate of $8.1\times10^{-11}\ M_\odot\,{\rm yr^{-1}}$
(assuming a distance of 10~kpc) suggests that the surface magnetic field
strength is below $\sim5\times10^{11}$~G. The 
corresponding cyclotron energy is thus below 4~keV, 
consistent with the non-detection of resonance features at high energies.
Although we could not unambiguously identify the optical counterpart, the
lack of a bright IR candidate within the $1'$ \xte\/ error circle rules
out a supergiant mass donor.
The neutron star in \src\ probably accretes from the wind of a
main-sequence O-B star, like most other high-mass binary X-ray pulsars.
The almost circular orbit
($e=0.036$) confirms the system's membership in a growing class of wide,
low-eccentricity systems in which the neutron stars may have received much
smaller kicks as a result of their natal supernova explosions.
\end{abstract}


\section{Introduction}

More than half of the known X-ray pulsars are transient sources which were
discovered during bright outbursts.  Most of these are Be-star/X-ray
binaries, which are believed to be progenitors of double neutron star
binaries \cite[e.g.][]{bhatt91}. In contrast to binaries with low-mass
($\la1\ M_\odot$)
companions, the neutron stars in these high-mass X-ray binaries (HMXBs)
typically accrete through the companion's stellar wind. The mass transfer rate
$\dot{M}$ --- as well as its variability --- depends in a sensitive manner
on the wind properties, as well as on the neutron star spin period and
magnetic field strength. A smaller class of pulsars accrete from
supergiant companions, which may fill their Roche lobes and thus accrete
persistently. Tidal forces in these binaries act to circularize the orbits
on a time scale much shorter than the active lifetime, while the wider,
wind-accreting binaries typically have moderate to high eccentricities
\cite[e.g.][]{bil97}. 
In recent years, a third class of binaries has emerged, with wide
$\ga20$~d orbits but low eccentricities $e\la0.1$.  These sources present
difficulties for the commonly accepted formation scenario in which the
natal supernova event imparts a ``kick'' to the neutron star, leading to
an initially eccentric orbit.  The tidal forces which act so efficiently
in the Roche-lobe filling systems cannot circularize wider orbits within
the source lifetime, suggesting that the initial kicks in
these wide, circular binaries may be unusually small \cite[]{pfahl02}.

For some transients the interval between outbursts can be as long as 20~yr.
Little is generally known about sources with such long duty cycles, due to
the dearth of observations (in particular with large-area modern
instruments with good timing capabilities).
For 
some candidate X-ray pulsars no pulsations have even been detected,
and the classification 
comes from a hard X-ray spectrum, typical in confirmed pulsars.  
\src\ ($l=37\fdg16$, $b=-1\fdg25$) is such a source, previously detected
just once before in outburst by {\it Uhuru}\/ and {\it Vela~5B}\/ in 1970--1
\cite[]{ftj76,pt84b}. Due to the
hard spectrum measured during those observations, the source was
tentatively identified as an HXMB. Consequently, we selected this source
as one of a group of hard transients with positions known to $10\arcmin$
or better, as candidates for target-of-opportunity observations by the {\it
Rossi X-ray Timing Explorer}\/ ({\it RXTE}).  In 2003 February a new
outburst of \src\ was detected by the All-Sky Monitor (ASM) aboard
{\it RXTE}\/ \cite[]{gal03a}.
The source was also detected with the IBIS and JEM-X hard X-ray
instruments aboard {\it INTEGRAL}\/ between 2003 March 10 and April 13
2003 \cite[]{mlg03}.
An \xte\/ Proportional Counter Array (PCA) scan across the {\it Uhuru}\/
position led to more precise coordinates of
$R.A. = 19^{\mathrm h}03^{\mathrm m}37\fs1$,
decl. = $+3\arcdeg11\arcmin31\arcsec$ (J2000.0),
with an estimated 90\% confidence uncertainty of $1'$ \cite[]{gal03c}.
Followup pointed \xte\/ 
observations detected coherent pulsations with a period of 2.763~s. 
Variations in the observed pulse frequency were also observed,
suggesting an orbital period of around 25~d.

Here we present timing and spectral analyses of \xte\/ observations
of \src\ throughout the 2003 outburst, as well as results from a search
for the optical/IR counterpart.

\section{Observations}

We made observations of \src\ with the Proportional Counter Array
\cite[PCA;][]{xte96} and the High-Energy X-ray Timing Experiment
\cite[HEXTE;][]{hexte96} instruments aboard \xte.
The PCA consists of 5 Proportional Counter Units (PCUs) each with a
collecting area of $\sim 1400\ {\rm cm^2}$ and a $1\arcdeg$ field of view,
that are sensitive to X-ray photons with energies in the range
2.5--90~keV.  Photon arrival times are measured to $\approx1\ \mu{\rm s}$,
while spectra are accumulated in up to 256 energy channels.
The HEXTE comprises two clusters, each with 4 scintillation detectors
sensitive to photons in the range 15--250~keV, collimated to view a common
$1\arcdeg$ field.  The detectors in the two clusters provide a total
collecting area of $1600\ {\rm cm^2}$.

Short ($\approx3$~ks) observations were scheduled every 3--4 days between 
2003 February 10 and July 16 (MJD~52,680 and 52,837)
in order
to adequately sample the 25~d candidate orbital period. In addition,
several longer observations were scheduled near the peak of the outburst
as part of a separate proposal to search for cyclotron resonance features
(PI: Heindl).
Data were analysed using {\sc lheasoft} release 
5.3
(2003 November 17).
We extracted PCA and HEXTE spectra from intervals within each observation
during which the center of the field-of-view was within $0.02\arcdeg$ of
the position of \src, and for which the limb of the Earth was more than
$10\arcdeg$ from the source direction. Spectra were extracted from
standard observing mode data for each instrument (``Standard-2'', with 129
channels between 2--60~keV for the PCA, and ``Archive'', with 64 channels
between 15--250~keV for the HEXTE).  PCA spectra were accumulated
separately for each PCU, and
instrument response matrices were generated for each PCU and each
observation using {\sc pcarsp} 
v.10.1.
We estimated background count spectra for the PCA using
``bright'' source models (suitable for when the count rate exceeds
$\approx40\ {\rm counts\,s^{-1}\,PCU^{-1}}$) developed for gain epoch 5
(2000 May 13 onwards) with {\sc pcabackest}.
We measured the mean flux for each observation by fitting spectra from
individual PCUs separately (using the model described in \S\ref{sec1})
between 2.5--25~keV\footnote{The Crab flux in this energy range is
$3.3\times10^{-8}\ \epcs$}.
We corrected the integrated 2.5--25~keV flux (except for PCU 2)
by dividing by the mean ratio of the
fluxes for each PCU relative to PCU 2,
and adopted the residual standard deviation on the rescaled fluxes as the
$1\sigma$ uncertainty.
For our timing analysis we used 
full-range PCA lightcurves with 4~ms time resolution, with time bins
corrected to the solar system barycenter.

\section{Results}

\subsection{Flux evolution and X-ray spectrum}
\label{sec1}

The X-ray flux peaked at almost $8\times10^{-9}\ \epcs$ (2.5--25~keV;
approximately 240~mCrab) around
2003 February 20 (MJD~52,690) and decreased linearly down to 
$\la10^{-10}\ \epcs$ by 2003 July 15 (Fig. \ref{fluxhist}).
Although bright outbursts in transient pulsars are sometimes
followed by extended periods of lower-level activity \cite[e.g.
KS~1947+300;][]{gal04c}, the ASM did not detect \src\ at a significant
 \centerline{\epsfxsize=8.5cm\epsfbox{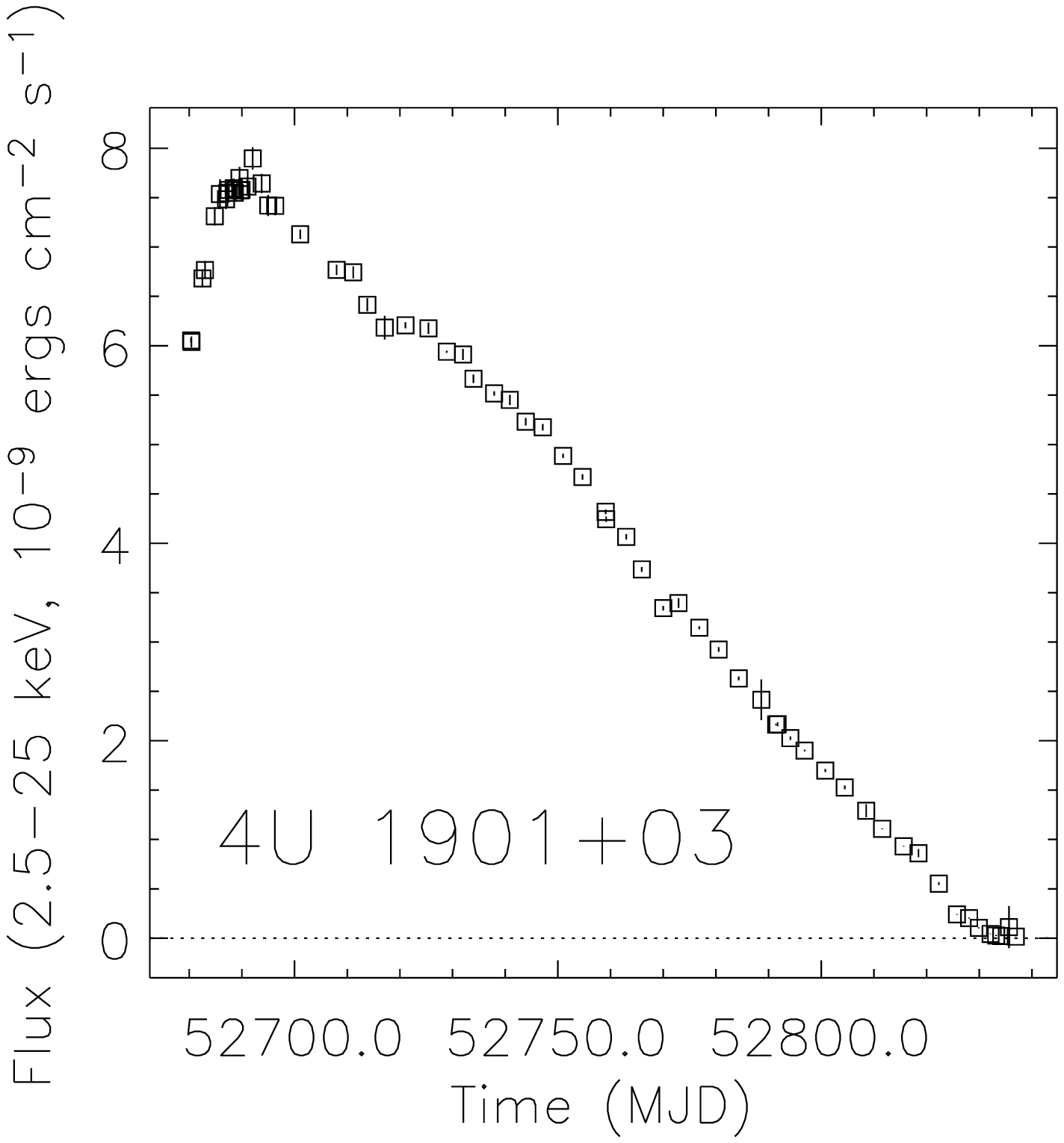}}
 \figcaption[]{X-ray intensity of \src\ throughout the 2003 outburst, as
measured by the \xte/PCA. Error bars indicate the $1\sigma$ uncertainties.
 \label{fluxhist} }
\noindent level between the end of the 2003 outburst
through 2005 June.  The modest PCA energy resolution 
at low energies meant that it was not
possible to constrain the column density $n_H$ for neutral absorption in
the spectral fits.
Thus, for all our spectral fits we froze $n_H$ at 
$1.2\times10^{22}\ {\rm cm}^{-2}$
(at the lower end of the expected range inferred from the $A_V$
estimates; see \S\ref{opt}). Our broadband (absorbed) flux measurements
were not sensitive to the assumed value of $n_H$.
Fits with commonly-used pulsar X-ray spectral models including power law,
cutoff power law and a combination of blackbody and power law gave
$\chi^2$ values indicating statistically unacceptable fits. We found the
best agreement with the data for 
a model consisting of a Comptonisation component
\cite[{\tt comptt} in {\sc xspec};][]{tit94} and a Gaussian component
centered around 6.4~keV to
represent fluorescent Fe line emission, both attenuated by neutral
absorption with column density at the survey value.
We assumed a systematic error of 1\%
in order to achieve a reduced $\chi^2$ (averaged over the fits to spectra
from each PCU) of $\approx1$.
A systematic error of this magnitude is typically required for PCA spectral
fits to bright sources, for example the Crab pulsar (R. Remillard, pers.
comm.).
Near the end of the outburst (between MJD~52,780 and 52,820) the
reduced-$\chi^2$ was somewhat larger,
between 2 and 3.5. The main factor contributing to the poor
$\chi^2$ value was a deficit of photons (compared to the model) between 8
and 10~keV; this deficit was observed irrespective of the
choice of continuum components tested.
In low signal-to-noise spectra (e.g. from short observations), the
residuals could be removed by including a blackbody component with
$kT_{\rm bb}\approx1$~keV, but for longer observations the residuals were
more complex.

Although the residuals indicate that the adopted spectral model does not
completely describe the source spectrum, the derived parameters present a
qualitative description that is adequate for flux measurements as well as
a general characterisation of the spectral shape and its
 \centerline{\epsfxsize=8.5cm\epsfbox{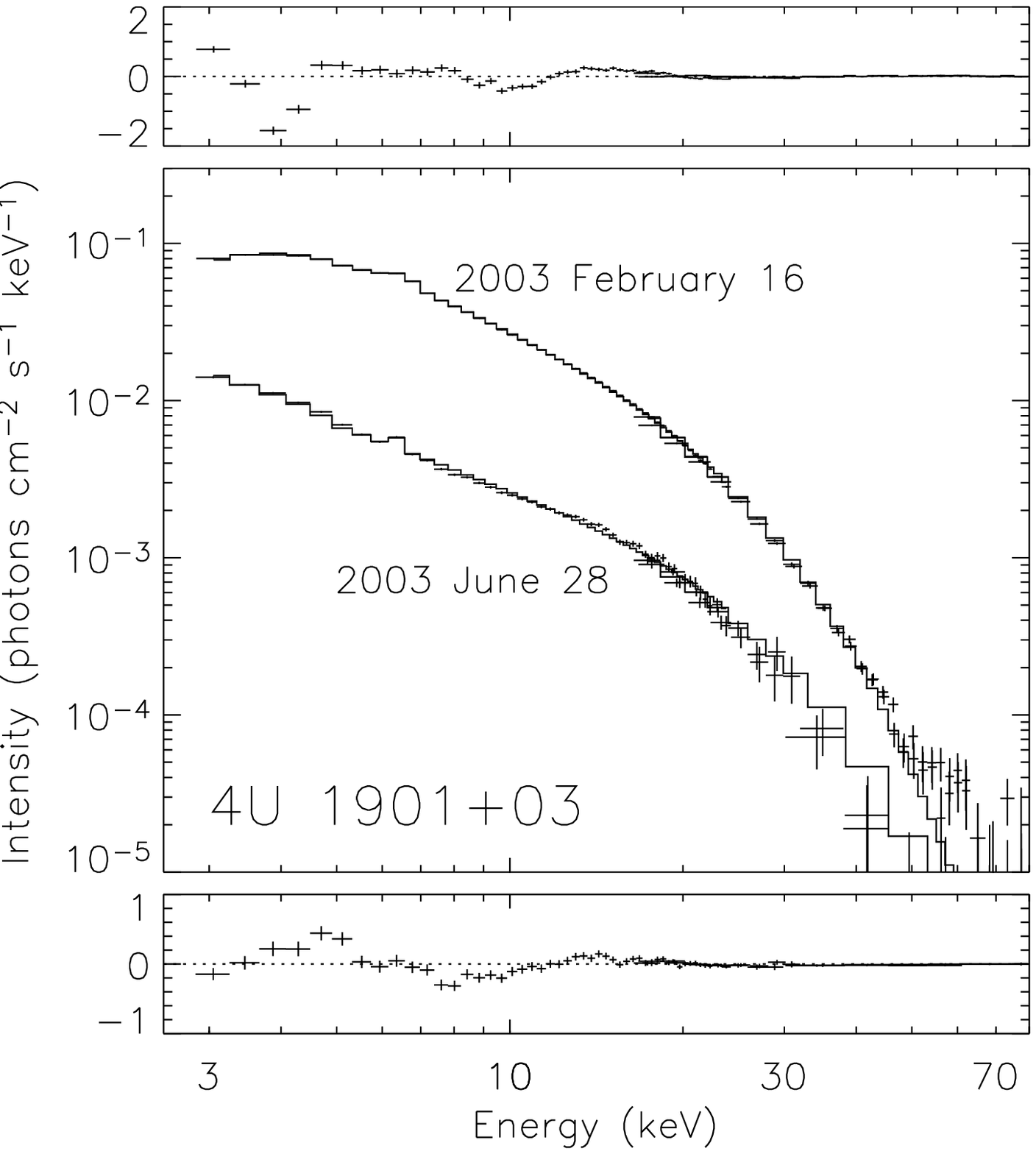}}
 \figcaption[]{Representative photon spectra of \src\ from the
peak (February 16, or MJD~52,686) and late in the decay (June 28,
MJD~52,818) of the 2003 outburst. The effect of the instrumental
responses have been removed (``unfolded''). Data from PCU \#2 are shown,
along with data from HEXTE clusters 1 and 2. For the 2003 June 28
observation, we rebinned the HEXTE spectra by a factor of 4. Error bars
indicate the $1\sigma$ uncertainties.
The top and bottom panels show the residuals to each model fit (units of
normalized counts$\,{\rm s^{-1}\ keV^{-1}}$) for the 2003 February 16 and
2003 June 28 spectra, respectively.
 \label{spectra} }
\vspace{0.25cm}
\noindent variation.
The fitted optical depth $\tau$ decreased slightly over the course of the
outburst, from 6 at the peak to 4 near the end of the outburst. The
temperature of the scattering electrons $kT_e$ increased over the same
period from 4 to 7~keV. A narrow emission feature around 
6.4~keV was
present throughout, with equivalent width between 60 and 150~eV.

We also made combined fits to PCA and HEXTE spectra for selected
observations near the peak and end of the outburst. The spectral
parameters for these broadband fits were similar to the fits to
the PCA data only. 
The spectrum was rather soft for a HMXB pulsar, with little emission
detected above 80~keV. As indicated by the evolution of the Comptonisation
model parameters, the spectrum hardened considerably over the course
of the outburst (Fig. \ref{spectra}).
Although the best-fit residuals still indicated
systematic deviations at energies $\la10$~keV from the best-fit model
spectrum, we found no evidence for cyclotron resonance features in the
spectrum.
Using the broadband fits, we estimated the bolometric correction as the
ratio of fluxes integrated over an idealised response matrix spanning
0.1--200~keV, and the flux in the range 2.5--25~keV, as 1.12.

\subsection{Pulse timing}
\label{pulse}

We estimated the pulse frequency for each observation by first folding the
4-ms lightcurve on a trial period, to obtain an observation-averaged pulse
profile with (typically) 32 phase bins. We then folded individually 
256-s segments on the same period, and cross-correlated the resulting
pulse profiles with the observation-averaged profile to obtain the phase
delay for each segment.  We then adjusted the period and repeated the
procedure until the phase delay exhibited no net trend with time
throughout the observation. The error was estimated from the uncertainty
on the first-order term of a linear fit to the phase delays. The resulting
frequency history shows approximately sinusoidal variations indicative of
Doppler shifts from binary orbital motion, superimposed on a significant
(non-linear) spin-up trend over the course of the outburst (Fig.
\ref{freqhist}).

We fit the frequency measurements with a linear model comprising the
spin-up due to accretion torques in addition to the apparent changes due
to orbital Doppler shifts:
\begin{eqnarray}
f(t) & = & f_{spin}(t)
   -\frac{2\pi f_0 a_X \sin i}{P_{\rm orb}} \nonumber \\
 & & \times (\cos l+g \sin 2l+h \cos 2l)
 \label{orbeq}
\end{eqnarray}
where $f_{spin}(t)$ is the time-dependent neutron-star spin frequency,
$f_0$ is a constant approximating $f_{spin}(t)$,
$a_X \sin i$ is
the projected orbital semimajor axis in units of light travel time,
and $P_{\rm orb}$ is the orbital period.  The coefficients $g\ (=e
\sin \omega)$ and $h\ (=e \cos \omega)$ are 
functions of the eccentricity $e$ and the longitude of periastron
$\omega$.  Finally, $l=2\pi(t-T_{\pi/2})/P_{\rm orb}+\pi/2$ is the
mean longitude, with $T_{\pi/2}$ the epoch at which the mean longitude
$l=\pi/2$.  For a circular orbit, $T_{\pi/2}$ is the epoch of superior
conjunction (when the neutron star is behind the companion).  The
right-most term in Eqn. 2 represents the orbital Doppler shifts to first
order in $e$; given the magnitudes of the uncertainties of our
measurements, this should be an adequate approximation as long as
$e\la0.2$.

We described the intrinsic spin frequency evolution with both $\dot{f}$ and
$\ddot{f}$ terms:
\begin{equation}
f_{spin}(t) = f_{0} + \dot{f}(t - t_{0}) + \ddot{f}(t - t_{0})^2
\label{torqeq}
\end{equation}
 \centerline{\epsfxsize=8.5cm\epsfbox{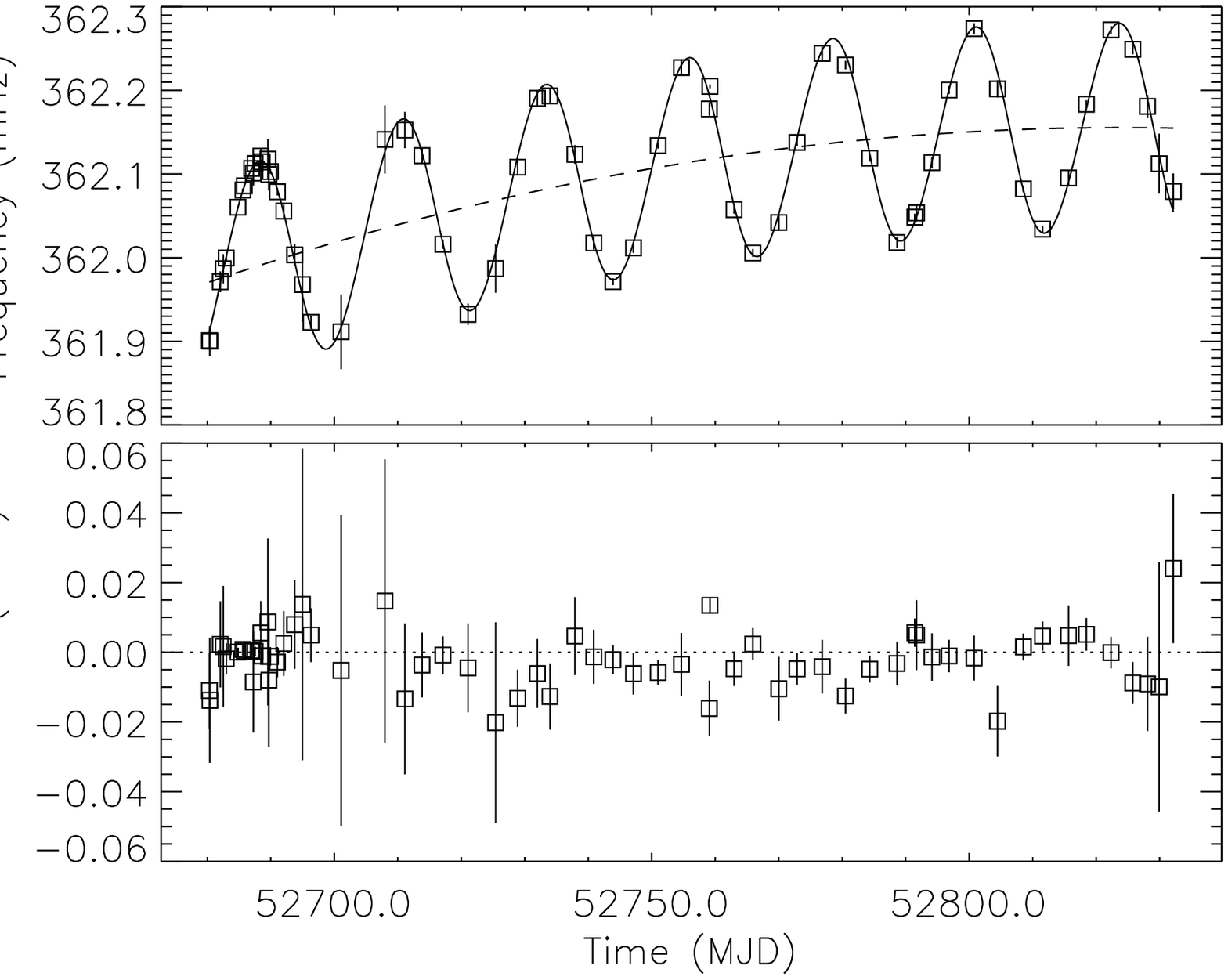}}
 \figcaption[]{Pulse frequency of \src\ throughout the 2003 outburst.
The top panel shows the measured frequency for each observation (open
squares) with error bars indicating the $1\sigma$ uncertainties. The
best-fit orbital model (Table \ref{orbit}) is plotted as a solid line; the
model for the intrinsic spin evolution of the pulsar (excluding the
orbital model) is plotted as a dashed line.
The lower panel shows the residuals to the orbital model.
 \label{freqhist} }
 \centerline{\epsfxsize=8.5cm\epsfbox{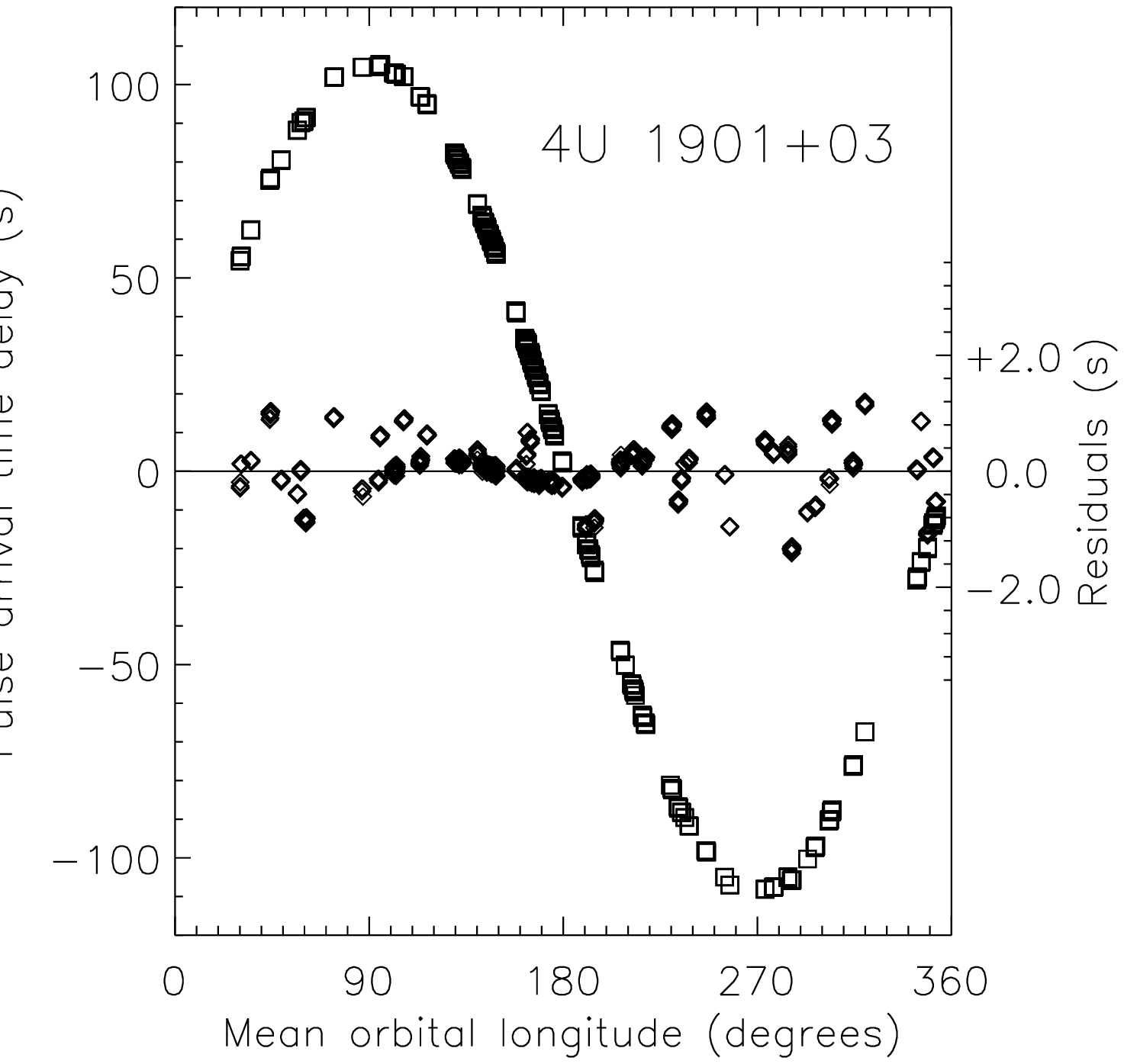}}
 \figcaption[]{Pulse arrival time delays due to the 22.6~d
binary orbit, with respect to a model for the intrinsic spin evolution of
the pulsar.  Squares indicate the measured time delays, the solid curve is
the best-fit orbital model, and the diamonds show the residuals from the
total model (on a $15\times$ expanded vertical scale). For a circular
orbit, the mean longitude is
the orbital phase angle measured from the ascending node; the pulsar is
behind the companion at $90\arcdeg$.
For such a low-eccentricity orbit, this remains approximately true.
 \label{deepto} }
\noindent where $f_{0}$ and $t_{0}$ are the frequency and time, respectively,
of the first frequency measurement and the $\dot{f}$, $\ddot{f}$ are
constant over the outburst duration.

We fit the frequency model to the measurements using a nonlinear
gradient-expansion algorithm ({\sc curvefit} 
in IDL). We achieved an
acceptable fit ($\chi^2=77.22$ for 53 degrees of freedom) with orbital
parameters 
$P_{\rm orb}=22.58$~d, $a_X \sin i=106.9$~lt-sec and $e=0.035$.
We used this preliminary orbital model to perform a 
fit of the
accumulated phase delay measurements. We first integrated
equations \ref{orbeq} and \ref{torqeq} to obtain an expression for the
pulse phase evolution with time:
\begin{eqnarray}
 \phi(t) & = & \phi_0+f_0(t-t_0)+\case{\dot{f}}{2}(t-t_0)^2+\case{\ddot{f}}{3}(t-t_0)^3 \nonumber \\
     & & -f_0 a_X\sin i\left(\sin l -\case{g}{2}\cos 2l -\case{h}{2}\sin 2l\right)
 \label{phaseq}
\end{eqnarray}
where $\phi_0$ is an arbitrary reference phase, corresponding in this case
to the peak of the fundamental of the first pulse observed in the first
observation. We then fit this model to the measured pulse arrival times,
defined as the peak of the fundamental Fourier component.

The frequency of observations during the first full orbital cycle of the
outburst (between MJD~52,680 and 52,700) was such that we were able to
unambiguously track the pulse phase over the entire cycle. Even so, we found
significant residual phase delays which varied systematically on a
timescale of a few days, with an rms amplitude of $\approx0.1$~cycles.
Although variations in the pulse profile over the entire outburst (see
\S\ref{sec2}) may contribute to the residuals to the phase fit,
residuals were also present during intervals when the pulse profile shape
was relatively consistent. Thus, there is significant intrinsic timing
noise present, perhaps arising from small changes in the instantaneous
accretion rate.

Beyond MJD~52700 the 2--3~d gaps between the \xte\/ observations introduced the
possibility of pulse count ambiguities, although only of magnitude
$\pm1$~cycle in general. We repeatedly computed the 
phase fit 
after adding or subtracting
a cycle within the data gaps, in order to minimise the total $\chi^2$
until no further improvement was possible.
Because of the timing noise, the resulting $\chi^2$ calculated using the
errors on the individual phase measurements
was much larger than the number of degrees of freedom.
In order to estimate the confidence limits for the orbital parameters, we
re-scaled the pulse arrival time errors so that the resulting $\chi^2$ 
was
1 per degree of freedom.
We then varied
each parameter in turn, fitting with all other parameters free to vary,
to determine the parameter range for which the rescaled $\chi^2\leq \chi^2_{\rm
min}(1+1/n)$, where $n$ is the number of degrees of freedom (1473 for the
full set of arrival time measurements). The resulting
orbital parameters and uncertainties are listed in Table \ref{orbit}; the
predicted frequency, intrinsic spin frequency for the neutron star and
the residuals from
the model are shown in Fig. \ref{freqhist}. The pulse arrival time delays
with respect to the intrinsic spin evolution model are shown in Fig.
\ref{deepto}.

Our best-fit parameters describing the intrinsic spin evolution indicate
that the spin-up rate was initially around $3\times10^{-11}\ {\rm
Hz\,s^{-1}}$, similar to the maximum measured for other transient pulsars
\cite[e.g.][]{bil97}.
According to the intrinsic spin model, the spin-up
decreased throughout the outburst, falling to zero just
before the end of the outburst, around MJD~52,824. The
 \centerline{\epsfxsize=8.5cm\epsfbox{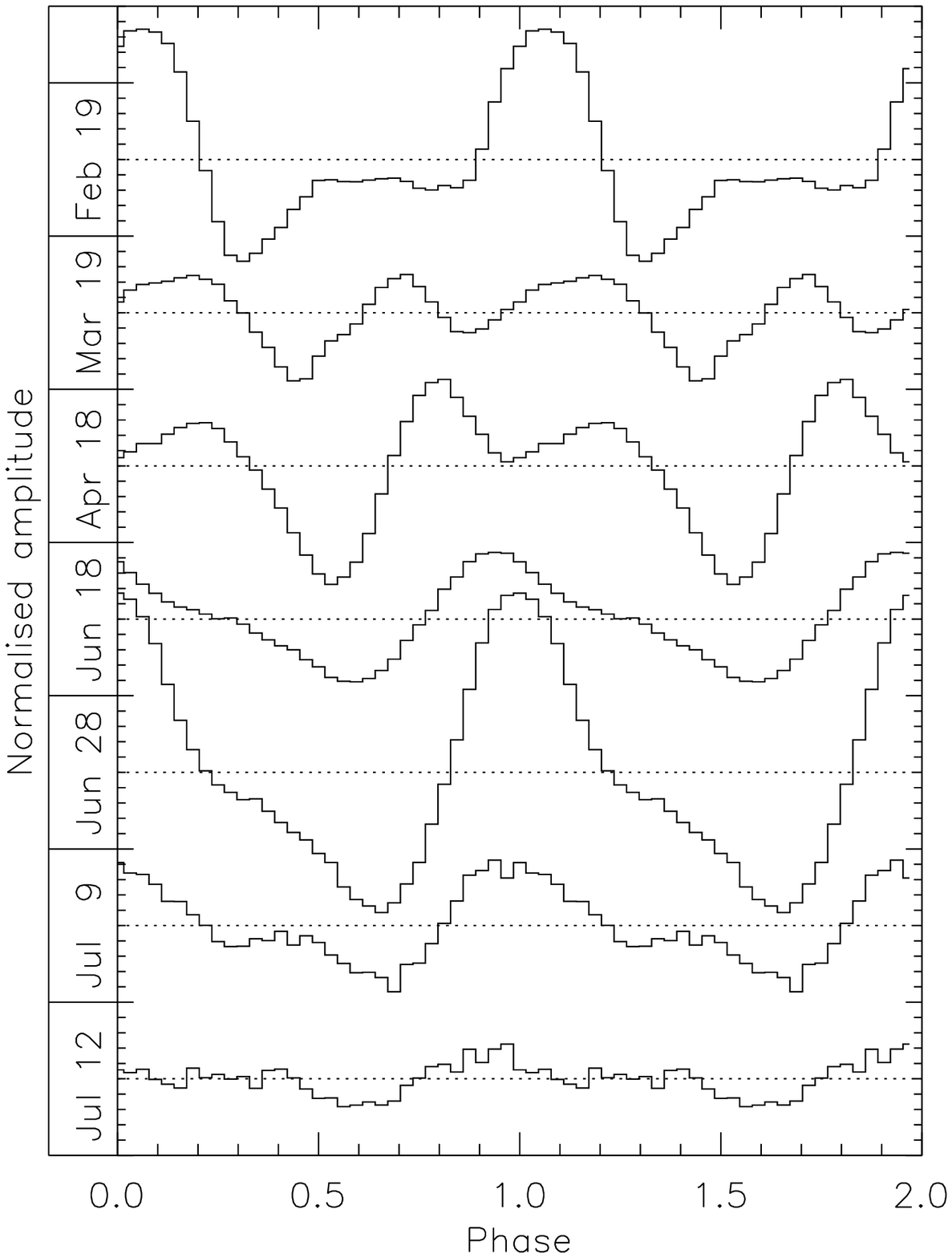}}
 \figcaption[]{Representative pulse profiles throughout the 2003 outburst
of \src. Phase 0.0 corresponds to the maximum of the fundamental. The
profiles have been rescaled by a factor of 3 to exaggerate the pulse
amplitude.
 \label{profiles} }
\noindent 2.5--25~keV flux by this time had
dropped to around $5.5\times10^{-10}\ \epcs$.

\subsection{Pulse profile variability}
\label{sec2}

The pulse profile was consistently non-sinusoidal, and exhibited several
intervals of stability punctuated by 
relatively rapid change.
The fractional pulse amplitude
also varied over the outburst, and was generally between
4 and 22\% (rms).
From the beginning of the outburst until 2003 March 3 the
profile was consistently similar to the example shown from February 19
(MJD~52,689; Fig.  \ref{profiles}).  
During that observation the
pulse amplitude was 15\% rms. Between March 3 and 13 the profile
switched to the characteristic double-peaked shape of March 19; also on
March 13 the 
rms amplitude dropped to 
7\%. The profile continued to evolve throughout March and April, with gradually
increasing pulse fraction and small drifts of the harmonic components
until another abrupt shift between 2003 June 14--18. The 
pulse amplitude had risen to 
13\% rms on June 14 before falling abruptly to 
9\% rms on June 18, and then recovering to an overall maximum for the
outburst of 
22\% rms on June 28.  After this final peak, the profile remained
similar in shape but with steadily decreasing amplitude towards the end of
the outburst.
Pulsations became undetectable ($<1$\% rms) after 2003 July 13
(MJD~52,833), by which time the 2.5--25~keV flux had dropped to below
$3\times10^{-11}\ \epcs$.

\subsection{A search for the optical/IR counterpart}
\label{opt}

Following the improved X-ray position obtained from the PCA scan
\cite[]{gal03c}, we examined the field of \src\ in Digital Sky Survey and
2MASS images to identify the optical counterpart.
The mass function 
implies a minimum companion mass (assuming a
$1.4M_\odot$ neutron star) of $4.5M_\odot$.
The most probable companion mass (for an inclination of $60^\circ$) is
$6.0\ M_\odot$.
The donor stars in long-period binary pulsars are typically either Be
stars or (sometimes Roche-lobe filling) OB supergiants.  
The column density interpolated from H{\sc i} survey observations towards
the source is between (1.1--$1.24)\times10^{22}\ {\rm cm}^{-2}$
\cite[]{dl90,bellHI}, which
translates to an $A_V=6.1$--6.9 \cite[assuming a standard dust to gas
ratio;][]{ps95}. 
Alternatively, the reddening estimated from dust IR emission is higher at
$A_V=10.2$ \cite[]{dust98}.
The USNO~A2.0 astrometric catalog \cite[]{usnoa2} contains just two stars
within the $1'$ error circle of \src\ having colors consistent with
early-type stars
suffering extinction with $A_V>6$ (i.e. $B-R\ga3$; Fig. \ref{field}, upper
panel).
We obtained low-resolution spectra of these two candidates
using the Low Dispersion Survey Spectrograph
(LDSS-2) on the 6.5~m Clay (Magellan II) telescope at Las Campanas, Chile.
We accumulated two 600~s spectra of each candidate on 2003
August 10, using the medium red grism with a $1\arcsec$ long slit, 
covering the range 
4500--9000~\AA. 
However, 
the overall spectral features
suggest that these
candidates are instead low-mass K0--5 stars (Fig. \ref{field}, lower panel), 
which are ruled out as the counterpart on the basis of the X-ray mass
function.
Clearly, it is not possible to distinguish between early-type stars with
high extinction and nearby late-type stars from optical photometry alone.

Thus, we also examined the $J$, $H$ and $K$ magnitudes of
candidates within $1'$ of the X-ray position from the 2MASS point source
catalogue\footnote{\url
http://www.ipac.caltech.edu/2mass/releases/allsky/doc/explsup.html}.
The expected range of
colors for a B-star 
for the estimated extinction range are
$J-H=0.5$--1.1 and $H-K=0.3$--0.6 \cite[]{asqu}. The colors expected for a
supergiant counterpart fall within similar ranges, slightly larger in
$J-H$ and smaller in $H-K$.
We found approximately ten stars in the $1'$ \xte\/ error circle with
colors within these ranges, including the optically identified
candidates A and B. Star B was the brightest in the IR bands, with
$J=11.5$; this is the only star consistent with the expected brightness of
a supergiant counterpart at $\la10$~kpc, and can be ruled out as the
counterpart on the basis of our spectroscopic observations. Two other stars
had $J$ magnitudes similar to that of star A, at $J\approx13$; we expect
these candidates are also nearby low-mass stars, like star A.
The remaining candidates all had $J>14$, much fainter than the limit of
$J\approx12$ expected for a supergiant companion at $d\la10$~kpc. Several
of these stars were not detected in the DSS image, suggesting that
$B\ga20$ perhaps
consistent with the upper end of the estimated $A_V$ range towards the
source. 
\vspace{0.5cm}
 \centerline{\epsfxsize=8.5cm\epsfbox{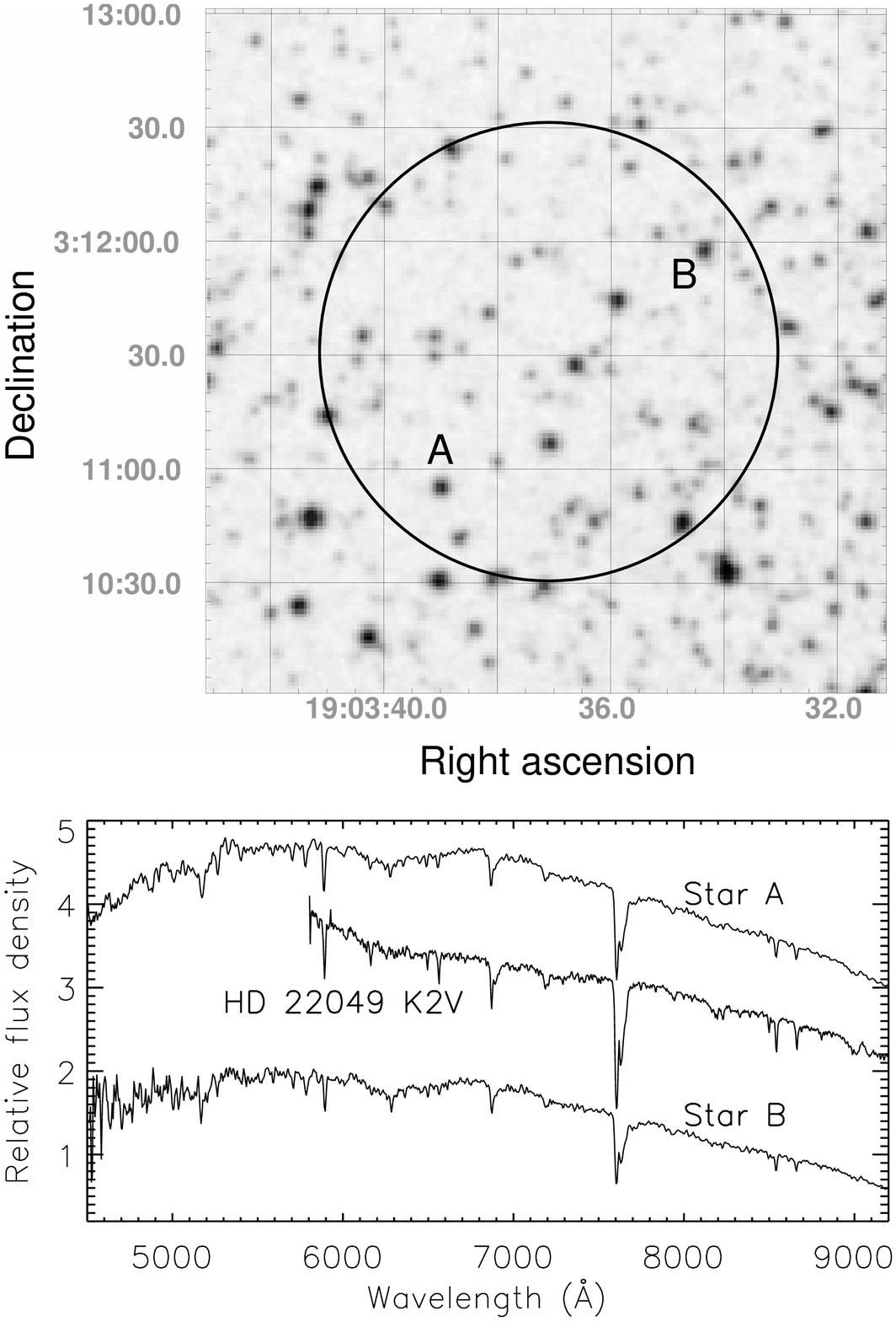}}
 \figcaption[]{{\it Upper panel}\/ Field of \src\ from the DSS POSS-II
data, showing the $1'$ {\it RXTE}\/ error circle and the two candidate
counterparts A and B identified by their large $B-R$ color indices. The
$B$-magnitudes 
measured from the USNO-A2.0 catalog, are 18.5 and 19.3, respectively; star
B is also the brightest star within the $1'$ error circle in IR at
$J=11.5$. 
{\it Lower panel} Magellan spectra of the candidates A and B, as well as a
comparison spectrum from a K2V star, HD~22049 \cite[]{dd94}.
The spectra of A and B were dereddened with $E(B-V)=1.1$ and 1.5 
respectively \cite[]{fitz99}.
The spectra of stars A and B are consistent with K0--5 stars and
thus can both be ruled out as the counterpart of \src.
 \label{field} }
\section{Discussion}
\label{disc}

\src\ is one of a small, but growing class of low-eccentricity
high-mass X-ray binaries.
Although we were unable to identify the optical counterpart,
the limit of $J\ga13$ for stars in the $1'$ \xte\/ error circle rules 
out a supergiant companion, unless the distance to the source is $>10$~kpc. 

Furthermore, the source is located on the Corbet diagram \cite[]{corbet86}
with confirmed Be transients such as 
4U~0115+63 \cite[e.g.][]{bil97}.
Like that source, \src\ exhibits infrequent outbursts that can span
multiple orbital periods, a behavior quite unlike the typically persistent
activity of partially Roche-lobe filling supergiants. 
If the mass donor 
is not filling it's Roche lobe, the efficacy of
tidal forces in circularizing such a wide orbit is negligible, which makes
the present low eccentricity of $0.036$ difficult to understand (given the
initially eccentric orbit expected to arise as a result of the natal
supernova kick).
As suggested by \cite{pfahl02}, the formation events for low-eccentricity
O-B transients like \src\ may be dynamically distinct from the more common
high-eccentricity binaries due to a much lower initial kick to the neutron
star.

The lack of eclipses indicates the inclination is $\la85\arcdeg$.
The 95\% upper limit on the
companion mass for an {\it a priori} isotropic distribution of inclination
angles is $88\ M_\odot$.
The estimated peak bolometric luminosity (for a distance of 10~kpc) was
$1.1\times10^{38}\ \eps$, and the integrated luminosity over the course of
the outburst was $7.4\times10^{44}(d/10\,{\rm kpc})^2\ {\rm ergs}$
(for a neutron star with $R=10$~km and $M=1.4\,M_\odot$).
The maximum intensity observed by {\it Uhuru}\/ during the 1970--1
outburst was $87\pm11\ \cts$ at epoch 1971.0, corresponding to
$1.5\times10^{-9}\ \epcs$ in the range 2--6~keV \cite[]{ftj76}.
Our broadband \xte\/ spectra suggest that 
20--30\%
of the source flux is
emitted in this energy range, so that the estimated maximum bolometric
luminosity for the {\it Uhuru}\/ observations was $\approx8\times10^{37}\ \eps$
(for $d=10$~kpc), consistent with the peak \xte\/ value to within the
error.
{\it Vela~5B}\/
measured a peak of $\approx8\ \cts$ around epoch 1970.9
\cite[]{pt84b}. The maximum observed flux was thus $3.6\times10^{-9}\
\epcs$ in the range 3--12~keV, in which range 60\% of the flux
observed by \xte\/ is emitted; thus, the estimated maximum bolometric
luminosity measured by {\it Vela~5B}\/ was $7\times10^{37}\ \epcs$, 
roughly consistent with both the {\it Uhuru}\/ measurements of the 1970--1
outburst and the the peak measured by \xte\/ for the 2003 outburst.
We also note that the estimated duration of the 1970--1 outburst
(excluding the peak around epoch 1970.6) was at least 120~d, similar to the
150~d duration of the 2003 outburst.
No emission prior to 2003 February was detected by the \xte/ASM, although
there is the possibility that one or more intervening outbursts occurred
sometime between 1971 and the launch of \xte\/ in 1995. A similarly bright
outburst
before 1980 would probably have been detected by {\it Vela~5B}\/ or the
ASM onboard {\it Ariel 5}\/ \cite[]{ariel5asm}. An outburst between 1987
and 1996 would likely have been detected with the {\it Ginga}\/ ASM
\cite[operational until 1991 November;][]{gingaasm} or the BATSE
experiment onboard {\it CGRO}\/ \cite[1991 April to 2000
June;][]{zhang95}.
Assuming no intermediate outbursts occurred,
we derive a time-averaged accretion rate for a 32.2~yr recurrence time of
$8.1\times10^{-11}(d/10\,{\rm kpc})^2\ M_\odot\,{\rm yr^{-1}}$.
We note that the 32.2~yr interval between outbursts in \src\ may be the
longest presently known 
for any
X-ray transient.

Whilst no cyclotron absorption features were detected in the X-ray
spectrum, the observed range of source flux over which pulsations were
detected allows a rough estimate of the dipole magnetic field strength of
the neutron star.
Assuming that a disk is present, it must be truncated above
the surface of the neutron star in order to allow the magnetic field to
channel the accreting material and produce observable pulsations. The
truncation radius $r_M$ is inversely proportional to the mass accretion
rate \cite[e.g.][]{fkr92}, so that the requirement for $r_M>R_*$ (where
$R_*\approx10$~km is the neutron star radius) even at the peak of the
outburst 
implies a lower limit on the magnetic field strength (although for 
\src\ this limit is several orders of magnitude below the canonical
field strength for long-period pulsars of $\sim10^{12}$~G).
For accretion to be dynamically feasible also requires that the
inner disk radius is within the corotation radius $r_{\rm co}$, i.e. the
radius at which the Keplerian orbital frequency equals the neutron star
spin frequency.
For the lowest flux at which pulsations were detected, this implies that
$B<0.5\times10^{12}(d/10{\rm kpc})^2$~G, giving a fundamental cyclotron
frequency of $\nu_{\rm cyc}\la4(d/10{\rm kpc})^2$~keV. This is consistent
with the absence of detectable cyclotron features in the broadband spectrum,
although it is possible that the residuals frequently present at
$\la10$~keV may arise from higher harmonics of a low-energy 
cyclotron absorption line. 
We can also estimate the magnetic moment by assuming that the long-term
mass accretion has left the pulsar close to spin equilibrium, i.e.
$r_M\approx r_{\rm co}$ \cite[e.g.][]{bil97}. In that case, we find
$B\approx 0.3\times10^{12}(d/10{\rm kpc})^{-6/7}$~G, consistent with the
above estimate.
We note that the inferred limit on
$\nu_{\rm cyc}$ and the low values of $kT_e$ are qualitatively consistent
with the observed correlation in other pulsars between the cyclotron
energy and the spectral cutoff \cite[e.g.][]{coburn02}.

Identification of the counterpart to \src\ is essential to 
confirm the nature of the mass donor suggested by these observations,
which may prove difficult unless the source becomes active again.
If the outbursts occur regularly every $\approx30$~yr, significantly more
advanced X-ray instruments may be available for the next outburst,
perhaps sufficient to measure the position more precisely and identify the
optical counterpart, as well as resolve the residuals below 10~keV and
measure the properties of the low-energy cyclotron lines, thus allowing
direct measurement of the field strength.

\acknowledgments

This research has made use of data obtained through the High Energy
Astrophysics Science Archive Research Center Online Service, provided by
the NASA/Goddard Space Flight Center.  
The Second Palomar Observatory Sky Survey (POSS-II) was made by the
California Institute of Technology with funds from the National Science
Foundation, the National Geographic Society, the Sloan Foundation, the
Samuel Oschin Foundation, and the Eastman Kodak Corporation.
This publication makes use of data products from the Two Micron All Sky
Survey, which is a joint project of the University of Massachusetts and
the Infrared Processing and Analysis Center/California Institute of
Technology, funded by the National Aeronautics and Space Administration
and the National Science Foundation.
This work was supported in part by
the NASA Long Term Space Astrophysics program under grant NAG 5-9184.


\begin{deluxetable}{lc}
\tablecaption{Orbital parameters from the pulse arrival time fit for 
  \src\label{orbit}}
\tablewidth{250pt}
\tablehead{ \colhead{Parameter} &  \colhead{Value\tablenotemark{a}} }
\startdata
$P_{\rm orb}$ (d)               & $22.5827 \pm 0.0002$ \\
$a_X \sin i$ (lt-s)       & $106.989 \pm 0.015$ \\
$T_{\pi/2}$ (MJD)               & $52682.2440 \pm 0.0010$ \\
$e$                             & $0.0363 \pm 0.0003$ \\
$\omega_{peri}$ ($\arcdeg$)     & $268.812 \pm 0.003$ \\
$f_{\rm X}(M)$ ($M_\sun$)       & $2.58348 \pm 0.00005$ \\
$t_0$ (MJD)                     & 52680.33163 \\
$f_0$ (mHz)                     & $361.970908 \pm 0.000009$ \\
$\dot{f}$ ($10^{-11}$~Hz\ s$^{-1}$) & $2.9759 \pm 0.0004$ \\ 
$\ddot{f}$ ($10^{-18}$~Hz\ s$^{-2}$) & $-1.1997 \pm 0.0003$ \\ 
 \enddata
\tablenotetext{a}{The uncertainties are at the $1\sigma$ confidence level,
i.e. the range of each parameter for which the rescaled $\chi^2 < \chi^2_{min}
+ \chi^2_{min}/$N(DoF).}

\end{deluxetable}

\end{document}